\documentclass[aps,prb,preprint,superscriptaddress]{revtex4-1}
\usepackage{natmove}
\usepackage[version=3]{mhchem}
\usepackage{graphicx}

\begin{document}

\title{Exciton Footprint of Self-assembled AlGaAs Quantum Dots in Core-Shell Nanowires}

\author{Yannik Fontana}
\thanks{Equal Contribution}
\affiliation{Laboratoire des Matériaux Semiconducteurs, Institut des Matériaux, Ecole Polytechnique Fédérale de Lausanne, 1015 Lausanne, Switzerland}

\author{Pierre Corfdir}
\thanks{Equal Contribution}
\affiliation{Cavendish Laboratory, University of Cambridge, J. J. Thomson Avenue, Cambridge CB3 0HE, United Kingdom}

\author{Barbara Van Hattem}
\affiliation{Cavendish Laboratory, University of Cambridge, J. J. Thomson Avenue, Cambridge CB3 0HE, United Kingdom}

\author{Eleonora Russo-Averchi}
\affiliation{Laboratoire des Matériaux Semiconducteurs, Institut des Matériaux, Ecole Polytechnique Fédérale de Lausanne, 1015 Lausanne, Switzerland}

\author{Martin Heiss}
\affiliation{Laboratoire des Matériaux Semiconducteurs, Institut des Matériaux, Ecole Polytechnique Fédérale de Lausanne, 1015 Lausanne, Switzerland}

\author{Samuel Sonderegger}
\affiliation{Attolight AG, EPFL innovation Square/PSE D, 1015 Lausanne, Switzerland}

\author{C. Magen}
\affiliation{Laboratorio de Microscopias Avanzadas
(LMA) - Instituto de Nanociencia de Aragon (INA) and Departamento de Fisica de la Materia Condensada, Universidad de Zaragoza, 50018 Zaragoza, Spain}

\author{Jordi Arbiol}
\affiliation{Instituci\'o Catalana de Recerca i Estudis Avan\c{c}ats (ICREA) and Institut de Ci\`encia de Materials de Barcelona, ICMAB-CSIC, E-08193 Bellaterra, Catalonia, Spain}
\author{Richard T. Phillips}
\affiliation{Cavendish Laboratory, University of Cambridge, J. J. Thomson Avenue, Cambridge CB3 0HE, United Kingdom}

\author{Anna Fontcuberta i Morral}
\email{anna.fontcuberta-morral@epfl.ch}
\affiliation{Laboratoire des Matériaux Semiconducteurs, Institut des Matériaux, Ecole Polytechnique Fédérale de Lausanne, 1015 Lausanne, Switzerland}

\date{\today}

\begin{abstract}
Quantum-dot-in-nanowire systems constitute building blocks for advanced photonics and sensing applications. The electronic symmetry of the emitters impacts their function capabilities. Here, we study the fine structure of gallium-rich quantum dots nested in the shell of \ce{GaAs}-\ce{Al_{0.51} Ga_{0.49} As} core-shell nanowires. We used optical spectroscopy to resolve the splitting resulting from the exchange terms and extract the main parameters of the emitters. Our results indicate that the quantum dots can host neutral as well as charged excitonic complexes and that the excitons exhibit a slightly elongated footprint, with the main axis tilted with respect to the long axis of the host nanowire. \ce{GaAs-Al_xGa_{1-x}As} emitters in a nanowire are particularly  promising for overcoming the limitations  set by strain in other systems, with the benefit of being integrated in a versatile photonic structure.
\end{abstract}

\maketitle

\section{Introduction}
The emergence of semiconductor nanowires (NWs) as a new class of functional material has triggered interest in the scientific community.  Several fields benefit from the opportunities brought by these nanostructures. Nanowires enable hybridization of fields, amongst which localized sensing\cite{Gao2010,Hahm2004}, electronic transport\cite{Bjork2004}, nanophotonics\cite{Grzela2012,Tribu2008,Claudon2010}, nanomechanics\cite{Yeo2014,Montinaro2014} and solid-state quantum optics\cite{Minot2007,Kouwen2010} to a degree hardly reached before. In particular, the latter aims at controlling and carrying quantum information with photons rather than electrons. In this context, NWs can provide significant and differential advantages. A workhorse in solid-state quantum optics is the system based on Stranski-Krastanov self-assembled quantum dots (QDs), usually grown on planar substrates\cite{Shields2007}. Despite excellent properties, in particular regarding linewidth and fine structure splitting, planar structures suffer from a poor light extraction efficiency mainly limited by total-internal reflection at the semiconductor/free-space interface. Serious efforts have been made to overcome this limitation, mainly through cavity engineering\cite{Pelton2002,Buckley2012}. In standing NWs, funneling the emitted light into well-defined modes --even non-resonant-- allows directional coupling to freespace. The read-out signal is improved significantly without the necessity of a radiation rate increase through the Purcell effect in a high-Q cavity. Outstanding results have been achieved with NWs in different geometries, illustrating the advantage of using NWs to mediate light-matter interactions\cite{Reimer2012,Munsch2013}. The bottom-up fabrication of QDs in NWs is usually achieved by modulating the composition of the semiconductor during the growth. A nanoscale region with a smaller bandgap acting as a QD is then defined\cite{Borgstrom2005}. Initially, the proximity of the QDs to the external surfaces and existence of crystal-phase mixing strongly limited the realization of narrow-linewidth emitters. Crystal-phase control as well as the ability to deposit in-situ an epitaxial protective shell resulted in impressive improvement of the inhomogenous broadening\cite{Dalacu2012,Yeo2011}. Nevertheless in this kind of QD the geometry is mostly determined by the NW core. Off-axis QD applications such as sensing or coupling to nanomechanical resonators are here precluded. 
Recently, small and localized Ga-rich islands nested in the AlGaAs shell of GaAs/AlGaAs core-shell NWs were identified\cite{Heiss2013}. These shell-QDs exhibit linewidth down to 30\,$\mu$eV and behave as bright single-photon emitters. High-resolution structural and chemical analysis on the QDs showed that they can form at the external part of the NW shell, making them ideal for sensing applications. The symmetry of electronic states of this new type of QDs has not been reported yet despite being an important parameter for a single-photons source. In this manuscript we therefore present a polarization and magnetic field dependent study of the light emission of these new type of QDs. We also show that the QDs can be loaded with extra carriers in addition to the primary electron-hole pair. Our results give important insights on the symmetry and localization properties of the excitons. This paper is structured as follows: in section \ref{methods} the sample and various measurement techniques used are described. The results are shown and discussed in section \ref{measurement}. Section \ref{conclusion} briefly sums up the results obtained on shell-QDs.

\section{Experimental details}\label{methods}
\subsection{Sample}
\begin{figure*}[htbp]
\includegraphics[width=8cm]{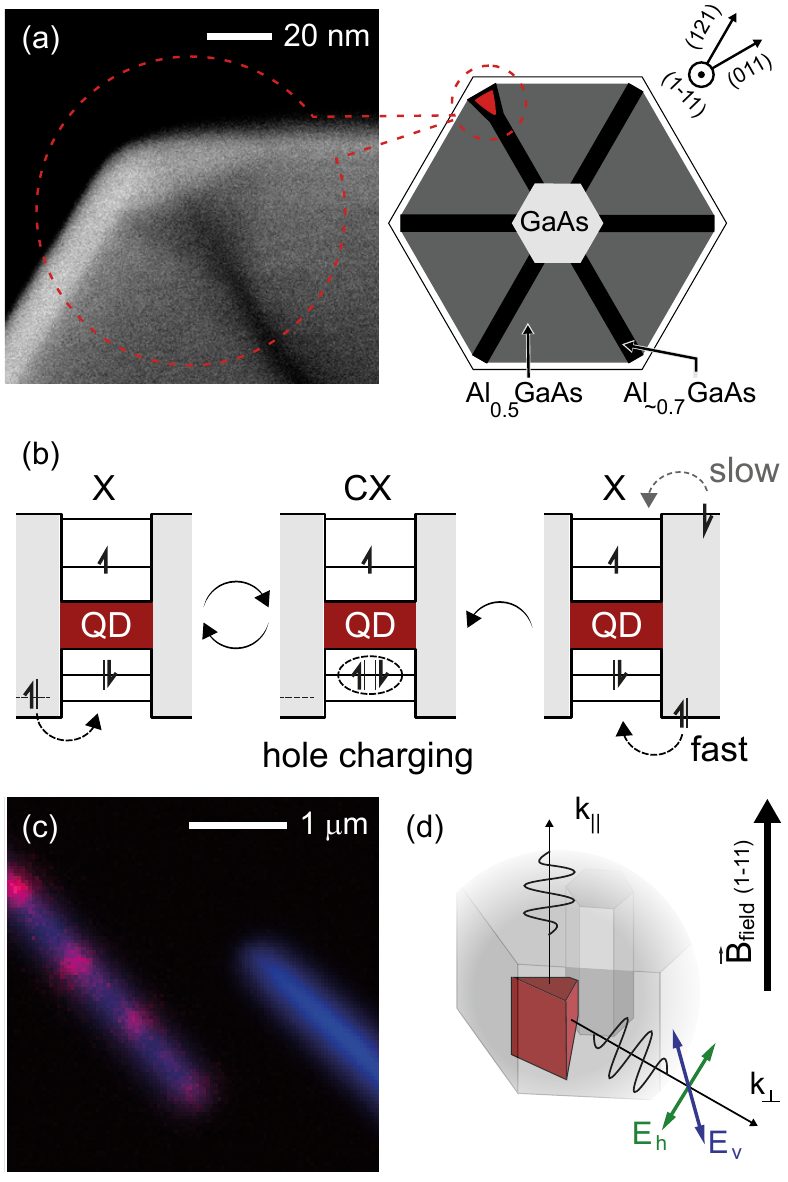}
\caption{(a) ADF STEM image of a segregated island nested in the AlGaAs shell of a NW. Darker contrast corresponds to higher aluminum fraction. The triangular feature corresponds to the Al-rich shell enclosing an Al-depleted island. A wide-scale drawing of such a cross-section is provided on the right; the space frame up-right indicates the important crystallographic directions. (b) Possible processes leading to the co-observation of exciton and charged-exciton (trion), here described for positive charging due to unintentional background doping (left) or an imbalance in free carrier capture rate (right). (c) 10\,K CL-SEM of two NWs. GaAs emission at 1.49-1.51 eV is color-coded in blue. Red is used to encode the signal recorded at 1.85 eV. The QDs luminescence at 1.85 eV is efficiently generated by the electron beam only very locally. (d) Tridimensional sketch of a segregated island embedded in the shell of a NW. The direction of the magnetic field used in this study is drawn, as well as the observation direction in the MPL experiment (k$_\parallel$ label). The observation direction for the polarization measurements done on NWs transferred on a substrate is also indicated (k$_\perp$ label).}
\label{figure1}
\end{figure*}

The shell-QDs structures studied here were grown by molecular beam epitaxy (MBE) on a DCA P600 system. GaAs NW cores were first obtained on a Si(111) substrate at 640\,$^\circ$C under a Ga flux equivalent to a planar growth rate of 0.03 nm/s and V/III flux ratio of 60, rotating the substrate holder at 7 rpm.\cite{Uccelli2011}. To grow the AlGaAs shells, the Ga flux was closed for about 5 min, the arsenic pressure was increased to 2$\cdot$10$^{-5}$ mbar and the substrate temperature decreased to 460\,$^\circ$C, thus switching the growth direction from axial to radial.\cite{Fontcuberta2008,Heigoldt2009}. The shell was 50 nm thick with a aluminum:gallium fraction of 51\%. The wires were further capped with a 5 nm GaAs protection layer to prevent oxidation. Figure \ref{figure1} presents the general characteristics of this type of NWs and QDs. 
Cross-sections of the nanowires perpendicular to the growth axis were prepared by mechanical polishing and ion milling. For annular dark field (ADF) scanning transmission electron microscopy analyses a TITAN 60-300 aberration corrected microscope operated at 300 keV was used.
The low temperature scanning electron microscopy (SEM) cathodoluminescence (CL) mapping were realized at 10\,K in a dedicated CL-SEM microscope (Attolight AG). The system allows  quantitative measurements thanks to a proprietary design. The light was collected and dispersed by a 300\,mm spectrometer and projected on an electron-multiplying charge-coupled device (EM-CCD).
The left panel of Fig. 1a shows an ADF STEM micrograph of the cross-section of the NW system. The global schematic of the cross-section of such NWs is presented in the right panel of Fig. 1a. The darker regions correspond to areas with higher Al-content. Regions with higher Al-content generally occur at the six vertices of the hexagonally shaped shell. The segregation process results from the different mobilities and sticking coefficients of Al and Ga on the facets and subfacets of the shell\cite{Skold2006}. This phenomenon is far from being trivial and depends on many factors, including the polarity of the \{112\} type subfacets\cite{Zheng2013}. During the growth, one of the Al-rich ridges may diverge and form a more complex structure where an Al-rich layer wraps around a Ga-rich island. In the ADF STEM cross-section of Fig. 1a, the NW is actually cut a few nanometers above the segregated island and allows the visualization of the segregated plane intersecting the Al-rich layer enclosing the Ga-rich island. Calculations confirmed that such islands may act as potential traps for electrons-hole pairs and thus behave as optically active QDs\cite{Heiss2013}. Indeed, sharp emission lines are observed in luminescence measurements. Figure 1b shows a schematic of the band structure of a generic QD and two examples of possible confined excitonic states (single and charged exciton, with different origins). Different charge states along with multiexcitonic states account for the observation of several emission peaks spaced by few meV.
We show in Fig. 1c a CL-SEM image revealing several emitters color-coded in red (emission ~1.85 eV). The bandgap emission at 1.51 eV is represented in blue and is fairly homogenous along the wire, while the red-encoded emitters are extremely localized. In the case of shell-QDs, multiple emitters can be found within the same NW, making the investigation of the effect of structural features on the QDs possible.

\subsection{Optical spectroscopy}
A drawback of both STEM and CL-SEM is the lack of information on the QDs symmetry, and on the electronic states. As schematically drawn in Fig. 1d, a typical way to obtain such information is to study the polarization-dependent emission properties of the QDs with polarization-resolved photoluminescence (PRPL). Magneto-photoluminescence (MPL) studies also provide further understanding on the symmetry and localization properties of the excitons.
For optical studies, the NWs were excited using the red-emitting, 632.8 nm line of a continuous wave Helium-Neon laser. For PRPL experiments, the samples were mounted in vacuum  on the cold-finger of a helium cryostat. The emitted light was first analysed by a Glan-Thompson polarizer followed by a half-waveplate, before being sent to a triple-stage spectrometer operating in additive mode. The light was detected by an electron-multiplying charge-coupled device (EMCCD).  For MPL, the sample was mounted at the bottom of the insert of a helium bath cryostat and kept at liquid helium temperature in a small He gas pressure. The sample was excited through the fiber-coupled objective mounted on the insert. The same objective was used to collect the PL in a confocal configuration, the single-mode optical fiber acting like a pinhole (more details in Ref. \citenum{Kehoe2010}). The PL was then dispersed on a single-stage 500 mm spectrometer and imaged on a CCD. The magnetic field was swept in 250 mT steps between 0 and 10 T.

\section{Measurements and discussion}\label{measurement}
Our detailed analysis of the QD electronic structure consists of an excitation power, polarization and magnetic field dependence microphotoluminescence study at the single NW and single QD level. This combination of methods allows the identification of the different emission peaks and gives information on the morphology of the shell-QDs, for which the application of typical 3D imaging techniques is highly challenging. We start with a general description of the electronic system. The ground state of an exciton (X) confined in a QD is usually composed of an electron and a heavy hole. The total angular momentum projection is $\pm 1/2$ for the conduction band electron and $\pm 3/2$ for the valence band heavy hole. We represent them further by their pseudo-spin value. In total four electron-hole combinations are possible: $\uparrow\Downarrow$, $\downarrow\Uparrow$, $\uparrow\Uparrow$, and $\downarrow\Downarrow$, where the single (double) arrow represents the pseudo-spin of the electron (hole). According to optical selection rules only the states with a composite total angular momentum M=$\pm 1$ are allowed to undergo radiative transitions: $\uparrow\Downarrow$, $\downarrow\Uparrow$\,. Allowed and forbidden transitions are usually referred respectively as bright and dark exciton transitions. If spin-related interactions between the electrons and holes are not considered, both the two bright and dark states are degenerate. The exchange interaction (EI) lifts the degeneracy between bright and dark states, which are split by a value $\delta_0$. Exchange terms also introduce a splitting of the forbidden doublet \emph{via} the so-called isotropic exchange interaction (IEI). The energy difference between the two dark states is labeled $\delta_2$ and is on the order of the $\mu$eV. In a QD with \ce{D_{2d}} or \ce{C_{3v}} symmetry, $\uparrow\Downarrow$, $\downarrow\Uparrow$ remain degenerate eigenstates of the system, exhibiting opposite circular polarization\cite{Bayer2002,Takagahara2000}. While symmetric QDs can be achieved by colloidal chemistry\cite{Moreels2011}, this is hardly the case for QDs in a semiconductor matrix. A myriad of causes such as shape, strain and/or alloy diffusion can introduce a slight asymmetry in the confining potential. The consequent anisotropic exchange interaction (AEI) hybridizes the bright states and split them by a value $\delta_1$. The two eigenstates are linearly polarized, which allows for their identification. Further, the hybridized eigenstates are mutually orthogonal. From now on, we refer to them as states H and V, from the laboratory reference frame. The splitting between H and V is representative of the anisotropy of the QDs and the polarization orientation reflects the elongation axis of the QD\cite{Young2006a}.
\subsection{Power-dependent microphotoluminescence}
In the excitation power and polarization dependence study, the NWs were dispersed on a Si substrate, in a lying configuration. We display in Fig. 2a a typical spectrum showing the emission of a single QD in a single nanowire. The broad peak at 1.51 eV corresponds to the emission from the GaAs core. The narrow lines in the 1.88 eV range are attributed to the QDs. It is important to note that the high brightness of the QD cannot be attributed to an enhanced carrier collection by the QD: from the CL-SEM scan in Fig. 2c, we can infer that the carrier capture length for the emitters is of the order of 170 nm (see supplementary figure S1). This low value is  a consequence of the GaAs core acting as a sink for carriers but also of the existence of a thin potential barrier around the QD. Figure 2b details the emission spectrum of a single QD in the 1.87 - 1.88 eV range. At the power used to acquire this spectrum (5\,$\mu$W), several emission lines can be observed. In order to understand the origin of the different observed peaks, we measured their luminescence intensity as a function of the optical power (P). The results are compiled in Fig. 2c. The intensities of the different peaks evolve differently with respect to P. This behavior makes us associate the peaks with different possible types of excitation confined in the QDs. The linear dependence with P of the intensity of the higher energy transition (black circles, labeled X) is characteristic of neutral exciton (X). In clear contrast, the intensity of the line at 1.871 eV exhibits a quadratic increase with P (orange squares), allowing us to ascribe it to biexciton (XX) recombination. The 6 meV XX binding energy of the shell-QD compares well to what has been reported for GaAs QDs made by droplet epitaxy\cite{Mano2009,Kuroda2013}. For other QDs, like InAs Stranski-Krastanov, the binding energy is usually slightly smaller. For InAs QDs in the strong confinement regime, the XX binding energy was found to be independent of the emission energy, while it increases for weakly confined GaAs natural QDs\cite{Moody2013}. Thus one may speculate that when GaAs QDs enter the strong confinement regime, the XX binding energy saturates as well; This would explain the consistent measurement of a XX binding energy around 4-6 meV for strongly confined GaAs QDs. The two lines at 1.874\,eV and 1.872\, eV show an intermediate behavior with P: as shown in Fig. 2b, their emission intensities are proportional to $P^{1.61}$ and $P^{1.53}$.  They correspond to two charged cases, where an exciton is accompanied by either an extra electron (negatively charged exciton, or trion X$^-$) or an extra hole (positively charged exciton, or trion X$^+$)\cite{Phillips1996}. Since we do not have the possibility to unambiguously differentiate X$^-$ and X$^+$, we will in the following simply refer to them as CX and CX' (red upward and blue downward triangles in Fig. 2c, respectively). Their origin can be manifold. Common mechanisms are a slight background doping in the semiconductor matrix as depicted in Fig. 1b (left), or a local imbalance in the injection rate of the optically generated carriers resulting from small electric fields or interface trapping of a prefered species (Fig. 1b right). The latter is likely to explain the CX peak, as background doping should be weak and the intensity would be expected to saturate at more moderate power.

\subsection{Polarization-resolved microphotoluminescence}
We now focus on a more detailed analysis of the X and CX emission by analyzing their polarization. For both X and CX, the emission polarization anisotropy is dominated by the NW-related antenna effect. Such an effect is commonly observed in NW-based systems where the NW diameter is smaller than (quasi-static case) or of the order of the wavelength (Mie-like resonances)\cite{Wang2001,VanWeert2009a}\, and is induced by the dielectric mismatch between the NW and its environment. We shown in Fig. 3a,b the CX and X polarization-resolved spectra. The CX emission line preserves the same spectral shape and shows a constant emission energy: variation in its intensity is the result of the antenna effect. In order to deconvolute the antenna effect from effects linked to the electronic structure, the spectra were thus normalized with respect to the CX peak. As it can be seen in Fig. 3b, the evolution of X emission spectra with polarization is much more intricate: the shape of the spectra varies with the polarization in a alternated pattern. We attribute this to the splitting of the bright exciton states induced by anisotropic exchange effect. To quantify the main orientation of the states H and V as well as the anisotropic exchange splitting energy, we performed a global fit on the whole dataset simultaneously. An example of a deconvoluted spectrum is shown in Fig. 3c. The spectra in Figs. 3a and b were fitted using two Gaussian functions to account for the broadening of the optical transition. Since we expect a similar broadening for the two peaks, their full-width at half maximum (FWHM) are constrained to equality. The latter point is justified by the relatively large broadening of the optical transition. If the main source of broadening was the radiative decay of X, one could expect different FWHM due to different dephasing rates of the split states\cite{Langbein2004}. Yet in this case the FWHM would reflect the lifetime of X, i.e. a linewidth around 200\,Mhz. The linewidths measured in our case are still orders of magnitude higher (in the Ghz range), and are most likely due to electrostatic fluctuation in the AlGaAs matrix, with a similar effect on both split states. The value obtained for the splitting $\delta_1$ between the two states H and V is 97\,$\mu$eV, as shown in Fig. 3c. If the H and V states FWMH are not constricted to equality, the extracted value for $\delta_1$ is of 107\,$\mu$eV, with a linewidth difference of around 5\%. This slightly higher value is obtained at the expense of a good convergence of the global fit; this is a consequence of the increase of fitting parameters, but can be considered as an upper limit. In comparison, droplet epitaxy GaAs QDs grown on (100) wafers showing geometrical anisotropy typically between 10 and 25\% exhibit an anisotropic splitting between 50 and 250\,$\mu$eV\cite{Abbarchi2008}.

We then show the polar plots for both  CX and normalized X. In the case of X the two hybridized eigenstates H and V are orthogonal and linearly polarized. Their main axes lie at 31.8\,$^{\circ}$ and 121.3\,$^{\circ}$ with respect to the NW long axis (Fig. 3d). As expected, the emission of CX does not show a fine structure splitting at zero magnetic field and is polarized along the NW axis (Fig. 3e, illustrating the strength of the purely photonic antenna effect). Figure 3f summarizes the characteristic polarization orientation of the CX peak and of the H and V states. The orthogonality between H and V states is expected and commonly observed in particular for systems with moderate or zero strain, like GaAs-AlGaAs, where the polarization axes are in-line and perpendicular with the elongation axis of the QD\cite{Young2006a,Abbarchi2010}. We emphasize that the polarization axes of the H and V states axis are not aligned with the NW axis. We can therefore conclude that the shell-QDs are not bound to be elongated along the NW axis, nor perfectly perpendicular to it.

\subsection{Magneto-photoluminescence}
Additional information on the shell-QDs symmetry was obtained using MPL. The experiments in magnetic field were carried out on as-grown nanowires, standing on the Si substrate, with the magnetic field parallel to the NWs axis. In this case, polarization could not be resolved. We display in Fig. 4 the evolution of X and CX emission energies when the magnetic field is increased from 0 to 10 T for two QDs, A and B. In the presence of a magnetic field, the peaks split and shift with magnetic field due to Zeeman and diamagnetic effects\cite{Kuther1998,Bayer2002,Besombes2000}\cite{Tischler2002}. The spectra of X and CX for magnetic fields between 0 and 10 T are plotted in Figs. 4b and 4d. The PL of both X and CX splits into four distinct lines. Based on their respective intensities at low magnetic field, these four transitions are attributed to bright and dark states. In fig. 4b and d, the peaks corresponding to dark states are indicated by arrows. Comparing QDs A and B, one can see that apart from slightly different splittings and shifts, the behavior is very similar. When the induced splitting are larger than the exchange terms, the dominant effects are given by the Zeeman contribution and by the diamagnetic shift, the energy of the peaks can be written as follows:
\begin{equation}\label{zeeman_equ}
E_{B/D}(B)=E^0_{B/D} \pm \frac{1}{2} \mu_B g^X_{B/D} B+\gamma B^2
\end{equation}
where $\mu_B$ is the Bohr magneton, $g^X$ and $\gamma$ are the exciton Land\'{e} factor and diamagnetic coefficient, respectively. The subscripts B and D refer to the bright and dark excitons,respectively. We assume $\gamma_B=\gamma_D=\gamma$ as the spin configuration should not affect significantly the diamagnetic coefficient\cite{Oberly2012}. Land\'{e} factors and diamagnetic coefficients can then be extracted by fitting the evolution of the X and CX emission lines as a function of B with Eq. \ref{zeeman_equ}. The fits allow one also to extrapolate the energy splitting between the bright and the dark states at zero field (shown in Fig 5a and c). Thus we can now attribute the CX-labeled peak to a charged exciton with certainty: only a spin-paired exciton complex can exhibit a vanishing IEI as can be seen in Fig. 4 and more clearly in Fig. 5a (red solid and dashed lines). For QD A, we obtain for the couple X/CX $\lvert g^X_{B/D} \rvert$ = 0.88/1.38, $\lvert g^{CX}_{B/D} \rvert$ = 0.45/1.44 , $\gamma_X$ = 5.8\,$\mu \textrm{eV}/\textrm{T}^{2}$,  $\gamma_{CX}$ =4.8\,$\mu \textrm{eV}/\textrm{T}^{2}$ and $\delta_0$ = 170\,$\mu$eV. Values for the Land\'{e} factors of other emitters are plotted in Fig. 5b as a function of the emission energy. Three main observations can be made: (i) In general in this energy range, the values between different QDs (given the same carriers and spins configurations) do not vary in an extreme way. (ii) a slight but noticeable trend can be seen: the Land\'{e} factors for all configurations tend to increase monotonically as the emission energy increases, as already reported in other systems\cite{Kleemans2009}. (iii) The Land\'{e} factors of the dark states are systematically larger than the bright ones. The $\delta_0$ value of our shell-QDs, with an average value of 249$\pm$ 54\,$\mu$eV is larger than what has been reported for the bulk value of GaAs\cite{Ishii2013}. This is expected for QDs excitons, as the isotropic exchange splitting energy $\delta_0$ can be seen as a measure of the QD volume\cite{Bayer2002}. While the $\delta_0$ values for our shell-QD is slightly smaller than what is usually reported for InAs lens-shape QDs\cite{Finley2002,Hattem2013}, it is similar or larger than the $\delta_0$ found for interface fluctuation GaAs QDs.

Coming to the X diamagnetic coefficient, it is directly related to the spatial extent of the X wavefunction in the plane perpendicular to the magnetic field:
	$\gamma=\frac{e^2}{8\mu_{\perp}}<\rho_{\perp}^2>$, 
where $\mu_{\perp}$ and $<\rho_{\perp}^2>$ are the exciton reduced mass  and the electron-hole correlation length in the plane perpendicular to the magnetic field. For a reduced mass of 0.068\,$m_0$ for \ce{Al_{0.1}Ga_{0.9}As} (corresponding to an intradot Al fraction of 0.1) the correlation length for QD A leads to an estimated confinement radius of about 8.3 nm and 7.5 nm for X and CX, in good agreement with the size of the segregated island measured by ADF STEM cross-sectional analysis in Fig. 1a. The smaller $\gamma$ value measured here compared to bulk or thick quantum well III-As systems is an indication of strong confinement, comparable with excitons confined in very localized fluctuation islands of narrow (2 nm) GaAs quantum wells or in droplet epitaxy QDs\cite{Erdmann2006,Abbarchi2010}. We underline also that the diamagnetic coefficient measured for CX is significantly smaller than the one of X. As discussed in Ref. \citenum{Hattem2013}, in the case of a negatively charged CX, this confirms that the QD is small and it indicates that the electron wavefunction is sensitive to the presence of a hole in the s-shell of the QD. In the initial state of the CX transition, the hole binds the electron to the QD. In contrast, after recombination of the CX, the remaining electron is less tightly bound to the QD ans its wavefunction spreads into the QD barriers, which results in the observed reduction of $\gamma$ for CX\cite{Fu2010,Schulhauser2002}. Technically, the latter argumentation holds for a positive (hole charged) CX as well. However due to the heavy mass of holes, the effect is bound to be weaker. In agreement with this, we attribute the observed variation in diamagnetic coefficient to the increased spreading of the additional carrier wavefunction in the QD barriers after recombination of CX. Considering the important (ca. 15\%) reduction between $\gamma_X$ and $\gamma_{CX}$, we further expect the CX emission line to correspond to the negative trion \ce{X^-}. Similar values of the diamagnetic coefficients can be extracted for other dots, giving an average value of $\gamma^{av.}_{X}$ =6.1$\pm$0.7\,$\mu \textrm{eV}/\textrm{T}^{2}$ and $\gamma^{av.}_{CX}$ =5.4$\pm$0.7\,$\mu \textrm{eV}/\textrm{T}^{2}$. 

We turn now to the discussion on the number of radiative transitions observed in the MPL measurements. For QDs with \ce{C_{2v}} symmetry only two optical branches are usually observed when the magnetic field is applied parallel to the QD high-symmetrxy axis. Any tilt between the magnetic field axis and the axis of high symmetry of the QD mixes the M=1 and M=2 states and allows the observation of four distinct optical transition\cite{Bayer2002,Witek2011,Hattem2013}. However the MPL properties of QDs with less common symmetries may differ significantly from what is observed for \ce{C_{2v}} QDs. First, QDs with low symmetry (approaching \ce{C_s}) can exhibit extra optically active states even at zero magnetic field \cite{Finley2002,Bayer2002}. QDs with \ce{C_{3v}} symmetry also show more than two emission lines when a magnetic field is applied parallel to the QD high-symmetry axis\cite{Karlsson2010,Sallen2011,Durnev2013}. In contrast, when the QD symmetry is elevated from \ce{C_{3v}} to \ce{D_{3h}}, only two emission lines are resolved in Faraday geometry\cite{Oberly2012}. Light and heavy holes mixing is also pointed out as being a cause of the observation of four emission lines. The mixing can occur \emph{via} strain\cite{Leger2007} or in case of an important elongation of the QD\cite{Besombes2000}.  Coming to QDs in NWs, recent works report the observation of two \cite{Witek2011,Corfdir2013}\, as well as four distinct emission lines \cite{Akopian2010}, when the magnetic field is parallel to the NW axis. As discussed in Refs. \citenum{Witek2011,Corfdir2013}, it is however not clear how interface roughness, surface, strain or crystallographic defects in the vicinity of the QD may affect the symmetry of the exciton states.
We discard symmetry elevation as the reason for the measurement of four emission lines since it corresponds to a very exceptional case. Strain is also ruled out as the GaAs/AlGaAs system, despite some recent investigation\cite{Hocevar2013}, is usually considered as being strain-free.
As shown in Fig. 5a, the upper branch of the X dark states comes into resonance with the lower branch of the bright states for a magnetic field of approximately 3 T. For QD with low or no symmetry, the mixing between bright and dark states results in an anti-crossing with an energy splitting scaling up with the symmetry breakdown. We could however not resolve any anti-crossing between the dark and the bright exciton states.Thus the anti-crossing magnitude is at least smaller than the exciton emission FWHM for this emitter (120\,$\mu$eV). In addition to the fact that no red-shifted peak associated with a strong symmetry breaking was measured at zero field\cite{Bayer2002,Finley2002}, the unability to observe an anti-crossing bewteen the dark and bright state allows us to conclude that the symmetry of the shell-QD is not severely degraded.

A natural explanation shows up if one considers results of the polarization-resolved experiment shown in Fig. 3d-f. The measurement revealed that the QDs do not have their elongation axis parallel to the NWs axis. In this case any magnetic field applied parallel to the NW axis mixes the bright and dark states of the QD. Therefore four emission lines instead of two can be observed. Indeed,  the larger the magnetic field, the larger the mixing between dark and bright states\cite{Witek2011}. This scenario is supported by the increase of the intensity ratio between dark and bright states with the magnetic field (Fig. 4b,d).

\section{Conclusions}\label{conclusion}
In conclusion, we described the fine structure of shell-QDs, a new declination of quantum emitters arising from segregation processes in the shell of GaAs/AlGaAs core-shell NWs. The analysis of the polarization and response to an external magnetic field shed light to key parameters of the QDs; based on these results, we could describe the typical morphology of the investigated shell-QDs: the emitter is asymmetric and its elongation axis does not coincide with the NW axis. We showed that charged excitons and biexciton  coexist with the neutral exciton. This is of particular relevance knowing that the exciton-biexciton cascade can be used to generate entangled photon pairs\cite{Benson2000,Akopian2006}, and that charged excitons can be harnessed to manipulate the spin of a single confined carrier or even nuclear spins\cite{Gerardot2008,Sallen2014}. The GaAs/AlGaAs combination is particularly favorable thanks to a vanishingly small lattice mismatch allowing the study of confined structures in an environement with minimal strain. Finally the inclusion of QDs inside NWs, beside enhancing the brightness also offers the possibility to couple the QD to a range of systems like nanomechanical oscillators, atomic emitters or even biological environments.

\begin{acknowledgments}
Financial support through the D-A-CH program of SNF (Grant nr 132506), EPSRC, the Hitachi Cambridge Laboratory and the ITN Nanowiring programs. ERC Starting Grant UpCon and NCCR-QSIT are greatly acknowledged.The research leading to these results has received funding from the European Union Seventh Framework Programme under Grant Agreement 312483 - ESTEEM2 (Integrated Infrastructure Initiative–I3).
\end{acknowledgments}

\bibliography{Excitonfootprintaps}

\begin{figure*}[htbp]
\includegraphics[width=8cm]{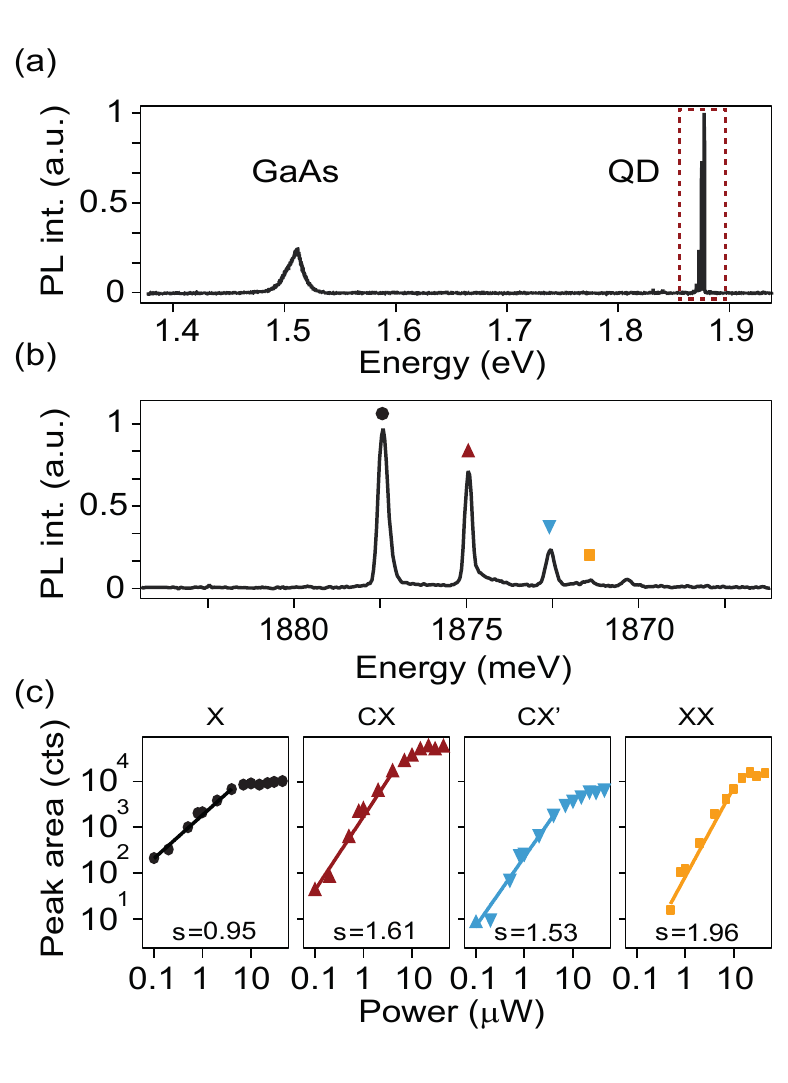}
\caption{Photoluminescence of an isolated shell-QD. (a) The emission intensities from both GaAs core and QD can be compared. The antenna effect expresses itself in a very bright emission from the nanoscale QD. (b) Spectral close-up of the QD outlined in (a). Several emission peaks can clearly be seen with an excitation power of 5\,$\mu$W (c) Evolution of the intensity of the four peaks marked in (b). The exciton (X, black circles) and biexciton (XX, orange squares) can be identified thanks to their linear and quadratic dependence on laser input power (fitted with a linear function of slope \textsf{s}). The two intermediate peaks show intermediate (superlinear) dependences, typical of charged excitons (CX and CX’, red and blue triangles).}
\label{figure2}
\end{figure*}

\begin{figure*}[htbp]
\includegraphics[width=7cm]{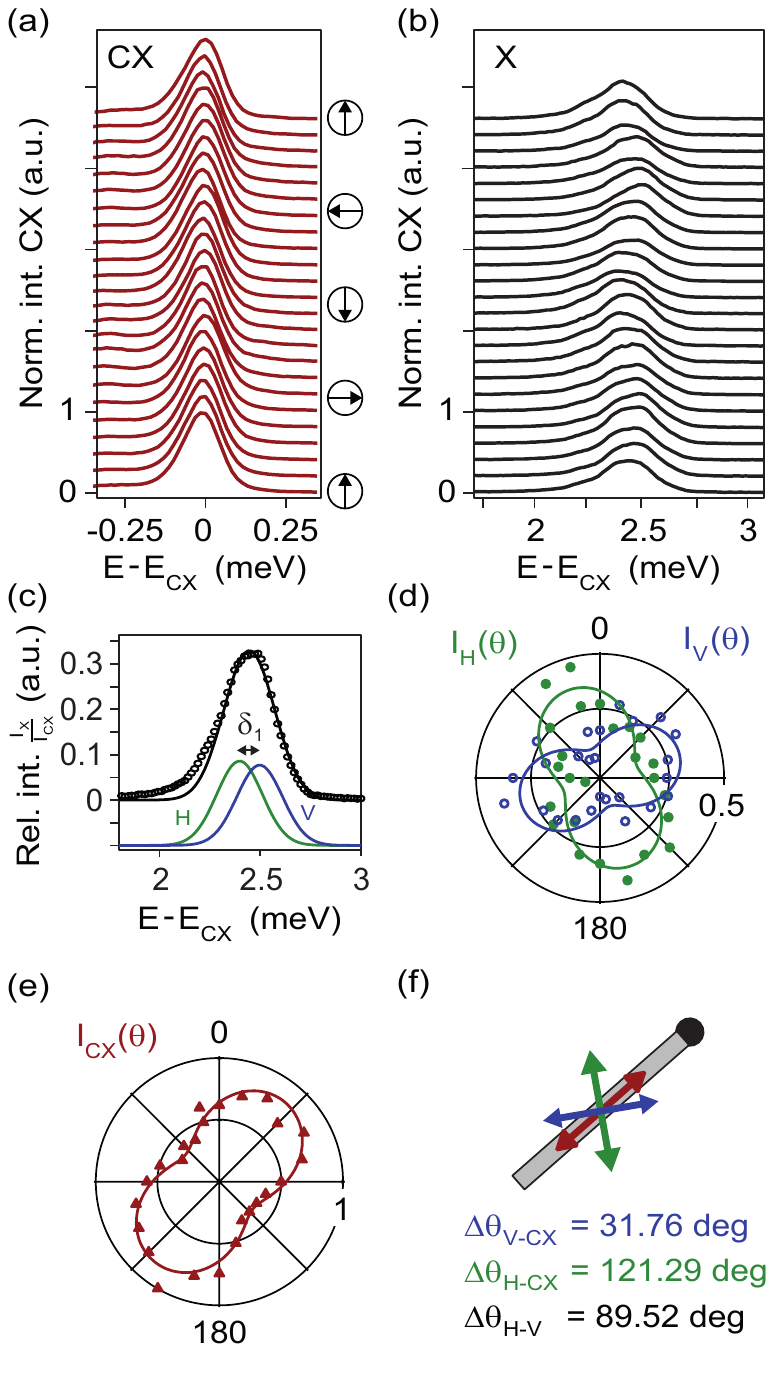}
\caption{Polarization analysis of X and CX emission lines. (a) Normalized emission from CX. The peak does not shift with polarization thanks to spin-pairing. The polarizer orientation in the lab reference frame is represented by the left-hand arrow quadrant (b) Emission from X. In contrast to its charged counterpart, the X emission peak shape varies as a function of the polarization as a result of the anisotropic exchange interaction between the hole and electron forming the exciton. (c) Example of an X spectrum fitted in order to retrieve the energy difference between the exchange interaction admixed states. Both linewidth and splitting result from a global fit on the full dataset. (d) Polar representation of the intensity of the two AEI-split states H and V. As expected, the states are orthogonal one to the other. (e) Intensity of the CX as a function of polarization. Due to spin-pairing, CX is protected against the exchange interaction. In this case, the intensity modulation comes from the photonic effect of the nanowire through the antenna effect. (f) Cartoon of the orientation of the different states. H and V states are orthogonal to each other, but neither of them coincide with the orientation given by the CX emission.}
\label{figure3}
\end{figure*}

\begin{figure*}[htbp]
\includegraphics[width=15cm]{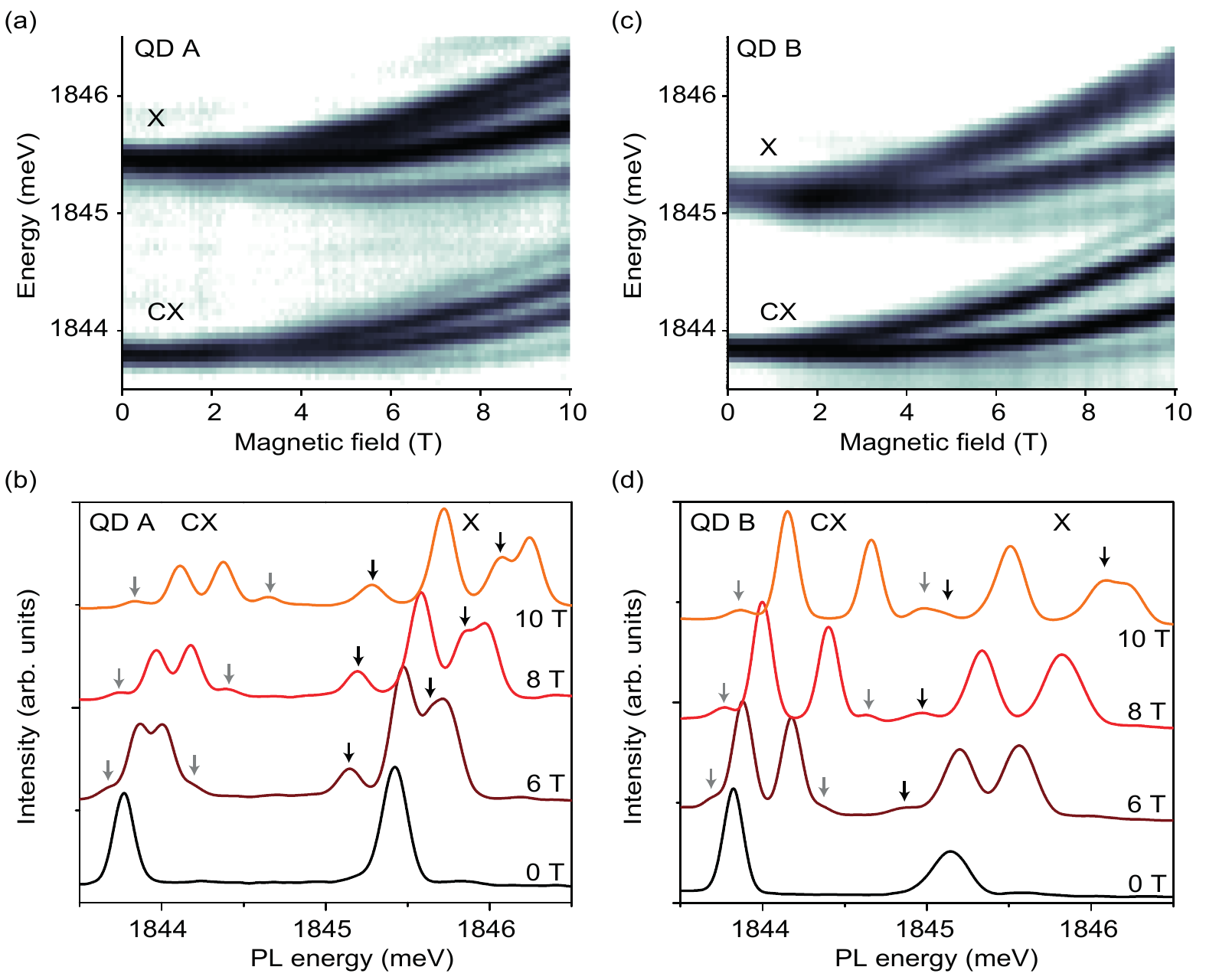}
\caption{MPL of two single QDs in a nanowire. (a,c) Color scans showing the evolution of the spectra of QD A (a) and QD B (c) as the magnetic field is progressively increased. In both case, X and CX are visible. The zero field peaks are split in groups of two doublets by the Zeeman effect and shifted because of diamagnetism. The bright and dark doublets are particularly obvious for the CX, as the splitting between dark and bright states at zero field is nonexistent. (b,d) Spectra at different magnetic fields for the QD A and B. the spectra are shifted in intensity for clarity. Black (grey) arrows denote the position of the dark peaks of the X (CX). The behavior of the dots is very similar, differing marginally  because of the different magnitude of the Land\'{e} factors, diamagnetic coefficients and X-CX binding energies.}
\label{figure4}
\end{figure*}

\begin{figure*}[htbp]
\includegraphics[width=15cm]{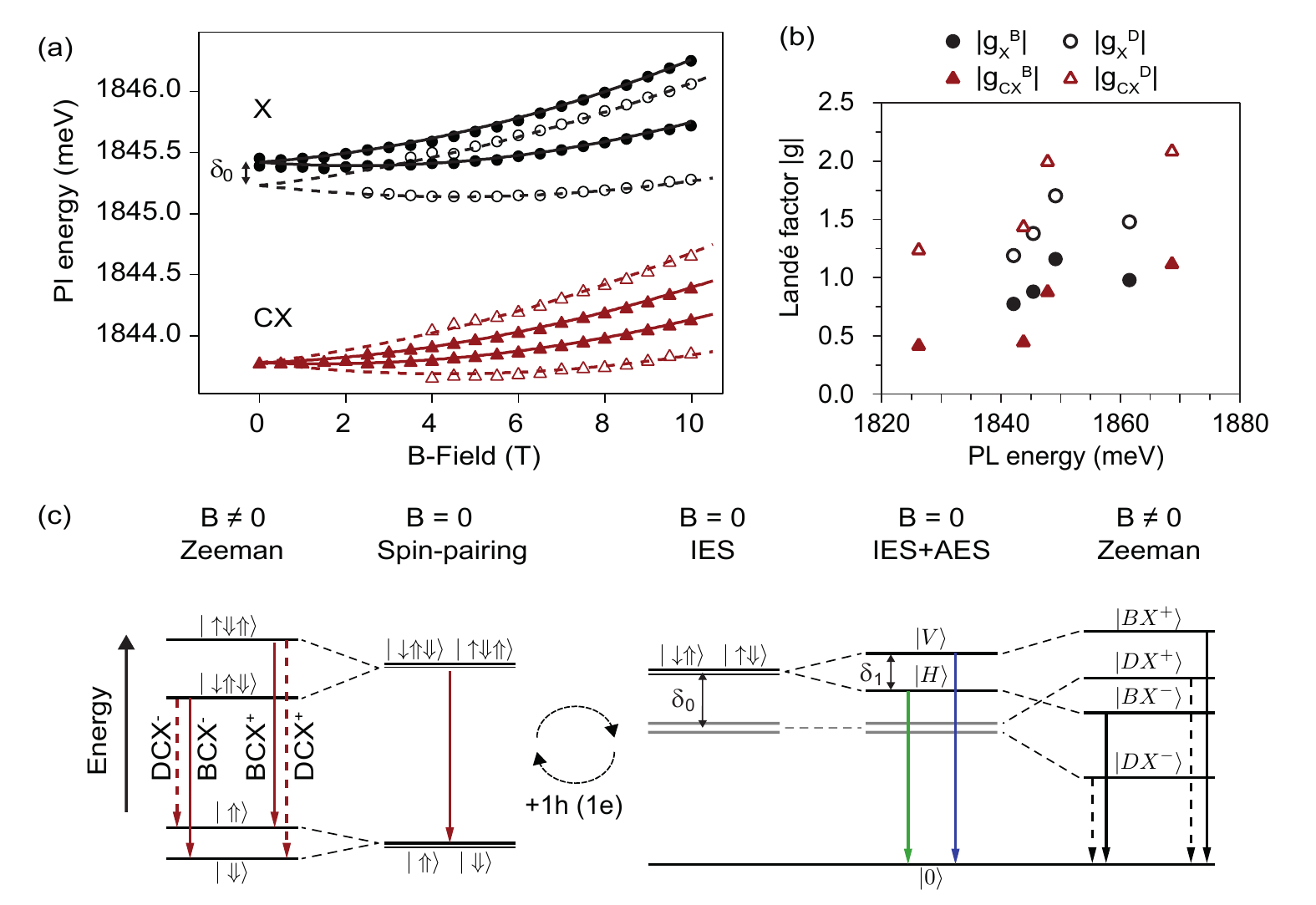}
\caption{(a) Fitting of the energetic splitting and shift for QD A with Eq. \ref{zeeman_equ}. (b) X and CX Land\'{e} factors from different QDs, showing consistent values between different QDs. The dark states $\lvert g_{D}^{X/CX}\rvert$ are systematical larger than the bright ones and the Land\'{e} factors of the bright and the dark states slightly increase with the emission energy. (c) Schematic depicting the splittings for X (right) and CX (left). For zero magnetic field, X states are split by the isotropic and anisotropic exchange interactions, while the spin-paired CX levels are still degenerated. The application of an external magnetic field lifts the degeneracy and mixes the pure states, leading to the observation of bright and dark optical transitions.}
\label{figure5}
\end{figure*}

\end{document}